%% file: proximity_and_similarity_october_2022.tex
\documentclass[12pt]{article}
\usepackage{array,multirow}
\usepackage{graphics}
\usepackage{color}
\usepackage{epigraph}
\usepackage[table]{xcolor}
\usepackage[dvips]{graphicx}
\usepackage{rotating}
\usepackage{array,multirow,graphicx}
\DeclareGraphicsExtensions{.bmp,.png,.pdf,.jpg,.tif,.wmf}
\usepackage{wrapfig,lipsum,booktabs}
\usepackage{bm}
\usepackage{enumerate}
\usepackage{lscape}
\usepackage{pdflscape}
\usepackage{array}
\usepackage{makecell}
\usepackage{rotating}
\usepackage[labelfont=sc]{caption}
\usepackage{verbatim}
\usepackage{multirow} 
\usepackage{hyperref}
\usepackage[utf8]{inputenc}
\usepackage{adjustbox}
\usepackage{footnote}
\usepackage{rotating}
\usepackage{geometry}
\usepackage{amsmath, amsthm, amssymb, latexsym}  
\usepackage{anysize}
\usepackage{tikz}
\usepackage{verbatim}
\usetikzlibrary{arrows,shapes}
\theoremstyle{plain}

\newtheorem{assumption}{Assumption}

\theoremstyle{definition}
\newtheorem{proposition}{Proposition}

\newtheorem{result}{Result}
\usepackage{sectsty}
\usepackage{natbib}
\usepackage{lipsum}
\usepackage{chngcntr}
\usetikzlibrary{decorations.pathreplacing}

\usepackage{caption}
\definecolor{dkgreen}{rgb}{0,0.4,0}
\definecolor{dkblue}{rgb}{0,0,0.4}
\definecolor{dkred}{rgb}{0.4,0,0}
\definecolor{mygrey}{rgb}{0.7, 0.7, 1.0}
\definecolor{myblue}{rgb}{0.1, 0.1, 1.0}
\definecolor{myred}{RGB}{0, 0, 0}
\definecolor{bluepic}{RGB}{0, 0, 255}
\definecolor{redpic}{RGB}{255, 0, 0}
\definecolor{greenpic}{RGB}{0, 255, 0}
\definecolor{indigo1}{RGB}{50, 100, 200}
\definecolor{indigo2}{RGB}{50, 100, 200}

\definecolor{g1}{RGB}{200, 200, 200}
\definecolor{g2}{RGB}{255, 0, 0}
\definecolor{g3}{RGB}{0, 255, 0}
\definecolor{g4}{RGB}{255, 0, 255}
\definecolor{g5}{RGB}{150, 150, 150}
\definecolor{g6}{RGB}{0, 0, 255}
\definecolor{g7}{RGB}{255, 255, 0}
\definecolor{g8}{RGB}{100, 100, 100}
\definecolor{g9}{RGB}{0, 255, 255}
\definecolor{g10}{RGB}{25, 25, 25}

\newcommand{\STAB}[1]{\begin{tabular}{@{}c@{}}#1\end{tabular}}

\tikzstyle{vertex}=[rectangle,fill=white,minimum size=10pt,inner sep=8pt]
\tikzstyle{types} = [circle, draw=black, fill=white,minimum size=25pt]
\tikzstyle{typep} = [circle, draw=black, fill=white, line width=1.2pt] 
\tikzstyle{group1} = [rectangle, fill=blue!24,minimum size=6pt]
\tikzstyle{group2} = [rectangle, fill=green!24,minimum size=6pt]
\tikzstyle{group3} = [rectangle, fill=red!24,minimum size=6pt]

\tikzstyle{edge} = [draw,thick,->]
\tikzstyle{weight} = [font=\small]
\tikzstyle{selected edge} = [draw,line width=1pt,-,red!50]
\tikzstyle{ignored edge} = [draw,line width=1pt,-,black!20]
\tikzstyle{group edge} = [draw,line width=5pt,-,blue!50]

\sectionfont{\large}
\hypersetup{
 colorlinks,
 citecolor=indigo2,
linkcolor=indigo2,
 urlcolor=indigo2
 }
 
 \newcommand*\mysize{%
   \@setfontsize\mysize{7.5}{9.0}%
}

\usepackage{tikz}

\definecolor{colortl1}{RGB}{56, 167, 207}
\definecolor{colortl2}{RGB}{109, 66, 176}
\definecolor{colortl3}{RGB}{250, 122, 122}

\begin{document}

\newgeometry{left=2.7cm, right=2.7cm, top=2.5cm, bottom=2.5cm}

\linespread{0.9}{
\title{Proximity, Similarity, and Friendship Formation:\\Theory and Evidence\thanks{The ``\textcircled{r}'' symbol indicates that the authors’ names are in certified random order, as described by \cite{ray2018certified}. We are grateful to Nicolás Fajardo and Linda Maokomatanda for their superb research assistance. We also thank Nicolás de Roux, Robert McMillan, Clementine Van Effenterre, Tatiana Velasco, David Zarruk, and participants at NEUDC and Toronto JAMS for valuable comments and suggestions. Funding for this project was generously provided by the Weiss Family Fund and the NAEd/Spencer dissertation fellowship. The experiment was approved by the MIT IRB (ID 1702862092), and is registered at the AEA RCT Registry (ID 0002600).}}

\author{A. Arda Gitmez\thanks{Department of Economics, Bilkent University. E-mail: \href{mailto:arda.gitmez@bilkent.edu.tr}{arda.gitmez@bilkent.edu.tr}.} ~ \textcircled{r} ~ Román Andrés Zárate\thanks{Department of Economics, University of Toronto. E-mail: \href{mailto:ra.zarate@utoronto.ca}{ra.zarate@utoronto.ca}.}}
\date{ \today }

\maketitle

\thispagestyle{empty}
\begin{abstract}
Can proximity make friendships more diverse? To address this question, we propose a learning-driven friendship formation model to study how proximity and similarity influence the likelihood of forming social connections. The model predicts that proximity affects more friendships between dissimilar than similar individuals, in opposition to a preference-driven version of the model. We use an experiment at selective boarding schools in Peru that generates random variation in the physical proximity between students to test these predictions. The empirical evidence is consistent with the learning model: while social networks exhibit homophily by academic achievement and poverty, proximity generates more diverse social connections.
\end{abstract}}

\clearpage
\pagenumbering{arabic} 

\linespread{1.2}

\section{Introduction}
\label{sec: intro}

The social connections that students form when they are young are known to affect them later in life \citep{chetty1_nature_2022, zimmerman}. This fact motivates one popular approach to improving social mobility in the form of school integration policies. School integration policies aim to help underprivileged students succeed by exposing them to peers from different backgrounds, thereby allowing them to make more diverse connections. Yet, as \cite{chetty2_nature_2022} suggest, increasing exposure at the school level does not necessarily result in more diverse friendships. This is due to {\it friending bias}: even when exposed to a diverse set of peers, students tend to form social ties with peers of the same race, income, or academic level \citep{chetty2_nature_2022, curr_jackson_pin_pnas, carrell}. The operation of this bias casts doubt on whether school integration policies can counter the segregating tendency in friendship networks, and actually succeed in bringing about more diverse friendships for underprivileged students. For instance, \cite{mele_paper} shows that race-based reallocation of students across schools does not necessarily decrease segregation by race in friendship networks.

As an alternative policy instrument, within-school policies offer a mean to tackle friending bias and foster diverse interactions. They do so by changing students' within-school exposure to specific peers through their allocation of classrooms, desks, and dormitories. Ultimately, within-school policies manipulate students' physical proximity, which is a central factor in forming friendships \citep{sacerdote_marmaros}. The overall effect of such policies depends on the underlying structure of the friending bias. If the friending bias is strong enough, proximity will not overcome homophily. Conversely, when proximity fosters friendships, within-school policies can improve the diversity of friendships by placing dissimilar students close to each other. 
 
Such a policy has a hidden cost, however.  As there are only a finite number of physically close locations, when school authorities place two dissimilar students together, they implicitly forgo putting two similar students together. If the effect of proximity on friendships for dissimilar pairs is positive but smaller than the effect for similar pairs, a within-school policy leads to a lower number of friendships (an efficiency loss). Understanding the link between friending bias and proximity is critical for designing policies that reduce segregation in friendship networks while not sacrificing the total number of friendships.

In this paper, we propose a model of friendship formation to explore the role of proximity and similarity in the diversity of friendships. We also provide an empirical test based on an experiment that randomly varies proximity between students at selective boarding schools in Peru. 

Our model is an application of the theory of exponential bandits \citep{keller/rady/cripps:2005}. In the model, a student decides whether to maintain interactions with a peer, where the value of maintaining the interaction is initially uncertain and is revealed over time. The optimal strategy prescribes an {\it exploration phase}, where a student engages with a peer to ``test the waters''. If she receives a signal that the interaction is valuable within the exploration phase, she maintains the interactions; otherwise, she stops engaging with that peer. The crucial assumption of the model is that when a student engages with a peer from a similar background, learning about the interaction's value occurs faster. Then, students end up being friends with similar peers more often than with dissimilar peers. That is, \textit{friendship patterns exhibit homophily}, which is a common empirical feature of social networks.

The model also explores the role of physical proximity. When a student engages with a peer in physical proximity, the cost of maintaining interactions is lower. A lower cost allows the student to explore the value of maintaining interactions longer, which translates into a higher friendship rate with proximate peers than with remote peers. That is, \textit{proximity fosters friendships}.

The model's distinguishing prediction is that the effect of physical proximity is stronger for dissimilar peers: \textit{proximity fosters diversity}. Intuitively, this is because proximity and similarity are substitutes in this model. When two peers have similar characteristics, the value of interaction is already explored even when the cost is high. In this case, a lower cost due to proximity has little impact on the outcome. By contrast, when two peers are dissimilar, the reduction in the interaction cost extends the exploration phase, allowing individuals to form new and more diverse friendships. The distinguishing prediction contrasts with an alternative version of the model we explore, where preferences are the main driver of homophily. In the alternative model, individuals derive a higher payoff from interacting with similar peers in expectation. In that case, the effect of proximity on friendships is higher for similar pairs, which is the opposite of our distinguishing prediction. 

We verify the theoretical predictions with an experiment at selective boarding schools in Peru. The details of the experiment are in \citet{zarate_jmp} that studies how more central and higher-achieving peers affect social and academic outcomes. Here, we exploit the fact that the peer assignment generates random variation in the physical proximity between a pair of students in the allocation to dormitories. Our empirical strategy leverages this variation to explore the effects of proximity on social interactions.

Our empirical analysis starts by exploring homophily in three dimensions: poverty status, baseline academic achievement, and baseline centrality level.\footnote{We use eigenvector centrality that measures a student’s influence within their social network. High values of this measure indicate that a student is connected to other individuals with high values of eigenvector centrality.} Our findings show that social networks exhibit homophily along these three characteristics. On average, poor students have 0.66 fewer non-poor friends (p-value $<0.01$) than non-poor students. Likewise, compared to higher-achieving students, lower-achieving students have, on average, 1.05 more interactions with other lower-achieving students and 1.26 fewer interactions with higher-achieving students. The results by other social skills variables (measured by centrality at baseline) also exhibit homophily. Compared to high central students, less central students have around 1.07 fewer connections (p-value $<0.01$) with more central peers. 

Next, we study the effect of proximity on social interactions. As part of the experiment, students' names were randomly organized on a list that school administrators used to assign students to specific beds and dormitories. Conditional on students' characteristics, the distance on the list is random, and it is a strong predictor of the physical distance between two students. We show that distance on the list is unrelated to social interactions before the intervention, as expected from this random variation. The effect of distance on social interactions after the intervention is high and statistically significant. Two adjacent students on the list are 16.6 percentage points more likely to become friends. This effect is 23.6 percentage points higher for first-year students, which is intuitive as they have no prior friendships with their peers. 

Finally, we explore the main theoretical prediction of the model: heterogeneous effects of proximity on social interactions by student characteristics. Our empirical findings are consistent with the predictions of the learning-driven homophily model. The impact of proximity on social connections is higher for students with different achievement levels (around 2.4 percentage points, p-value 0.028) and poverty statuses (around 3.4 percentage points, p-value $<0.01$ ). These results suggest that policies that increase within-school exposure across both characteristics can succeed in generating more diverse connections.

\textbf{Related literature.} This paper contributes to three branches of literature. 

First, we contribute to the literature on social networks in schools. While there has been extensive work estimating homophily by type in the formation of direct and indirect friendships \citep{curr_jackson_pin_pnas, mele_paper, depaula}, less is known about the effects of mixing students of different types within a school. We show that even though students prefer friends similar to them, these differences are less prevalent when students are physically close to each other. This finding is an essential factor to consider when designing policies for school integration. Our results suggest that within-school policies that mix students by poverty status or different academic achievement levels can generate more diverse social connections.  

Second, we contribute to the literature that examines how exposure to peers of a different type changes attitudes towards that group. When two students have different poverty statuses and academic achievement levels, the effect of exposure is stronger than when they have similar characteristics. These results add to the evidence of contact on attitudes and preferences such as race \citep{kremer}, poverty \citep{rao}, caste \citep{lowe}, and religion \citep{mousa}. Our results also align with those of \cite{gallen/wasserman:2022}, who show that while there is homophily by gender in the selection of mentors, this homophily disappears once information about mentors is provided to students. 

On the theoretical front, we propose a novel model of friendship formation. As in \cite{curr_jackson_pin_ecta}, our model also predicts homophily. In our model, faster learning (rather than preferences) is the driver of homophily. 
\cite{kets-sandroni:2019} also propose an information-driven theory, where homophily alleviates strategic uncertainty because similar individuals receive correlated impulses. In comparison, our model makes predictions on the effects of proximity on friendship patterns. \cite{peski:2008} and \cite{baccara-yariv:2013} propose models of homophily where group formation is a critical part of the argument; in contrast, our model considers the isolated instance of a student choosing whether to interact with a peer. 

\section{The Model}
\label{sec: model}

We model the friendship process as a game of {\it experimentation} with exponential bandits. The model is based on \cite{keller/rady/cripps:2005} and our treatment is closest to \cite{bardhi/guo/strulovici:2020}.\footnote{For other models of experimentation, see \cite{keller/rady:2010}, \cite{keller/rady:2015}, and \cite{strulovici:2010}.} The model allows us to decompose the effects of proximity (a {\it design} variable) and similarity (students' {\it exogenous} characteristics) on friendship patterns and understand their interaction. 

\subsection{The Friendship Formation Process} \label{section:process}

There is a finite set of students $I$. At any point in time, each student $i \in I$ decides whether to maintain interactions with the remaining students in $I \setminus \{i\}$. All the interactions between all pairs occur independently and simultaneously. In what follows, we describe interactions between a representative pair $(i,j) \in I \times \left(I\setminus \{i\}\right)$.

There is a finite set of categories $\mathcal{K}$. Each category $k \in \mathcal{K}$ represents a binary classification based on observable characteristics (poverty levels, academic achievement, etc.) Student $i$'s characteristic in category $k$ is given by:
\begin{align*}
    \tau_{ik} \in \{0,1\}
\end{align*}
For instance, when $k$ represents poverty status, $\tau_{ik} = 1$ means $i$ has high poverty, and $\tau_{ik} = 0$ means low poverty. Student $i$'s {\it type} is $\tau_i \equiv \{\tau_{ik}\}_{k\in \mathcal{K}}$.

Consider a pair of students $(i,j)$. Based on students' types, we can introduce a comparison notion for the pair's {\it similarity}. For any characteristic $k$, let:
 \begin{align*}
     \theta_{ijk} \equiv \begin{cases}
     1, & \quad \text{if } \tau_{ik} = \tau_{jk},\\
     0, & \quad \text{otherwise.}
     \end{cases}
 \end{align*}
We say that the pair $(i,j)$ is {\it more similar} than another pair $(i,\ell)$ if: $\theta_{ijk} \geq \theta_{i\ell k}$ for all $k \in \mathcal{K}$, with $\theta_{ijk} > \theta_{i\ell k}$ for some $k\in \mathcal{K}$.

After $i$ and $j$ observe $\tau_i$ and $\tau_j$, they start the following process of maintaining interactions with each other. The time is continuous, denoted by $t \in [0, \infty)$, and the discount rate is $r > 0$. At any $t$, $i$ chooses $a_{ijt} \in \{0,1\}$. Here, $a_{ijt} = 1$ if $i$ maintains interactions with $j$ and $a_{ijt} = 0$ if not. Simultaneously, $j$ maintains interactions with $i$ by choosing  $a_{ijt} \in \{0,1\}$.

Students $i$ and $j$ are at a physical distance $d_{ij} = d_{ji} \in \{\underline{d},\overline{d}\}$ from each other,\footnote{Because our empirical specification uses two possible distance levels ({\it neighbors} versus {\it non-neighbors}), we opted to present the theoretical model accordingly. The main insights would go through with a continuous measure.} with $\underline{d} < \overline{d}$. Maintaining the interactions imposes a flow cost of $c(d_{ij}) = c(d_{ji}) > 0$ on the student who maintains the interaction. We impose the following structure on the cost. 
\begin{assumption} \label{assumption:cost}
Proximity reduces the cost of interactions, i.e., $c(\underline{d}) < c(\overline{d})$.
\end{assumption}

For $i$, maintaining an interaction with $j$ may be {\it valuable} or not. When an interaction with $j$ is valuable for $i$, $i$ receives a flow payoff of one, and otherwise, $i$ receives a payoff of zero. The value is denoted by $v_{ij} \in \{0,1\}$. The realization of $v_{ij}$ is initially unknown to $i$, who has a prior $p_0 \equiv \Pr(v_{ij} = 1)$. Similarly, $j$'s value from interacting with $i$ is $v_{ji} \in \{0,1\}$, which is drawn independently with a prior $\Pr(v_{ji} = 1) = p_0$. We restrict attention to the case where $p_0 > c(\overline{d})$: each interaction is valuable in expectation, even when $j$ is far away.

Student $i$ can learn the value of interaction with $j$ via a signal that arrives at exponentially distributed random times with parameter $\lambda_{ij}$. In particular, the probability that $i$ receives a signal within the time interval $[t, t + dt)$ is 
\begin{align*}
    a_{ijt} \cdot v_{ij} \cdot \lambda_{ij} \cdot dt
\end{align*}
Because only valuable interactions generate signals, $i$ is {\it certain} that $v_{ij} = 1$ upon the arrival of the first signal. This is the moment when $i$ ``clicks'' with $j$ and realizes the interaction will be worth maintaining: it is a {\it breakthrough} as in \cite{keller/rady/cripps:2005}. The symmetric argument applies to student $j$, who receives her signals about $v_{ji}$ with rate $\lambda_{ji} = \lambda_{ij}$.

The arrival rate of signals depends on the types of $i$ and $j$. The crucial condition we impose on $\lambda_{ij}$ is as follows.
\begin{assumption} \label{assumption:learning}
Suppose the pair $(i,j)$ is more similar than another pair $(i,\ell)$. Then, $(i,j)$ learn their values of maintaining interactions faster than $(i,\ell)$, i.e., $\lambda_{ij} > \lambda_{i\ell}$.
\end{assumption}
Assumption \ref{assumption:learning} captures the fact that individuals from similar backgrounds have ease in communication, perhaps because they share some common knowledge \citep{mayhew:1995} or because they share the same mental models of the world \citep{carley/palmquist:1992}. 

All in all, $i$'s expected discounted payoff from interacting with $j$ using an action profile $\{a_{ijt}\}_{t \geq 0}$ is:
\begin{align*}
    \mathbb{E}\left[\int_{0}^{\infty}r e^{-rt}  a_{ijt} \cdot \left( v_{ij} - c(d_{ij})\right) dt \right]
\end{align*}
where the expectation is taken over $v_{ij}$, the stochastic arrival of signals, and $\{a_{ijt}\}_{t \geq t}$. The optimal strategy $\{a^*_{ijt}\}_{t \geq 0}$  maximizes the total expected payoff of student $i$ at each history $h_{ijt}$. Here, $h_{ijt}$ involves the action profile up to period $t$, $\{a_{ij\tilde{t}}\}_{\tilde{t}< t}$, and the arrival of $i$'s signals. Student $j$'s payoff and optimal strategy $\{a^*_{jit}\}_{t \geq 0}$ is defined similarly.

\subsection{Analysis} \label{section:analysis}

The following proposition characterizes the optimal strategy of student $i$ when interacting with another student $j$. Due to the independence of learning processes, the strategy of $j$ when interacting with $i$ is symmetric. The proposition is a corollary of Proposition 3.1 of \cite{keller/rady/cripps:2005} and Lemma A.1 of \cite{bardhi/guo/strulovici:2020}. For the sake of conciseness, we are omitting the proof here.

\begin{proposition} \label{proposition:optimalsolution}
In the optimal strategy, $a^*_{ijt} = 1$ if and only if the belief $p_{ijt} = \Pr(v_{ij} = 1 \vert h_{ijt})$ is above
\begin{align*}
    p^*_{ij} \equiv  \frac{r \cdot c(d_{ij})}{r + \lambda_{ij} - \lambda_{ij} \cdot c(d_{ij})}
\end{align*}
\end{proposition}

A couple of notes about the equilibrium dynamics are in order. First, consider the posterior belief at time $t$, $p_{ijt} = \Pr(v_{ij} = 1 \vert h_{ijt})$. Due to the {\it breakthrough} nature of the signal, the posterior decreases as the $i$ maintains interactions with $j$ and the signal does not arrive. As soon as the signal arrives, the posterior beliefs jump to $p_{ijt} = 1$.

Second, the cutoff belief and the rate of change of beliefs imply an ``exploration phase'' where $i$ {\it experiments} by waiting for a signal to arrive. If the signal arrives within the exploration phase, $i$ maintains interactions with $j$ from then on. If it does not, $i$ stops her interactions with $j$ onwards. By Lemma 3.1 of \cite{keller/rady/cripps:2005}, the length of the exploration phase is:
\begin{align*}
    t^*_{ij} & = \frac{1}{\lambda_{ij}} \log \left(\frac{p_0}{1-p_0} \frac{1-p^*_{ij}}{p^*_{ij}}\right)\\
& =\frac{1}{\lambda_{ij}} \log \left(\frac{p_0}{1-p_0} \frac{r+\lambda_{ij}}{r}\frac{1-c(d_{ij})}{c(d_{ij})}\right)
\end{align*}
$p_0 > c(\overline{d})$ ensures that $ t^*_{ij} > 0$, i.e., there is an exploration phase for each pair. 

The ex ante probability that $i$ maintains interactions with $j$ sufficiently far ahead in time is:
\begin{align*}
    \Pr\left( a^*_{ijt} = 1 \vert t \geq t^*_{ij} \right) & = p_0 \cdot \left( 1 - e^{-\lambda_{ij} \cdot t^*_{ij}} \right)\\
    & = p_0 \cdot \left( 1 - \frac{1-p_0}{p_0} \frac{r}{r+\lambda_{ij}}\frac{c(d_{ij})}{1-c(d_{ij})} \right)
\end{align*}
Similarly, the ex ante probability that $j$ maintains interactions with $i$ sufficiently far ahead in time is:
\begin{align*}
    \Pr\left( a^*_{jit} = 1 \vert t \geq t^*_{ji} \right) & = p_0 \cdot \left( 1 - \frac{1-p_0}{p_0} \frac{r}{r+\lambda_{ji}}\frac{c(d_{ji})}{1-c(d_{ji})} \right)\\
    & = p_0 \cdot \left( 1 - \frac{1-p_0}{p_0} \frac{r}{r+\lambda_{ij}}\frac{c(d_{ij})}{1-c(d_{ij})} \right)
\end{align*}
where the second equality uses $d_{ij} = d_{ji}$ and $\lambda_{ij} = \lambda_{ji}$.

Our empirical analysis is conducted in the dyad level, and we are interested in the likelihood that at least one student within the pair maintains interactions. Therefore, our outcome of interest is:
\begin{align*}
    y_{ij}(d_{ij}) \equiv \Pr\left(a^*_{ijt} = 1 \text{ or }a^*_{jit} = 1 \vert t \geq \max\{t^*_{ij}, t^*_{ji}\}\right)
\end{align*}
Given the analysis so far, this is equal to:
\begin{align} \label{eqn:yij}
    y_{ij}(d_{ij}) & = 1 - (1-p_0)^2 \left(1 + \frac{r}{r+\lambda_{ij}}\frac{c(d_{ij})}{1-c(d_{ij})} \right)^2 
\end{align}
It is important to emphasize that {\it learning} about $v_{ij}$ and $v_{ji}$ is the essential part of this equation. If $v_{ij}$ was revealed to $i$ and $v_{ji}$ was revealed to $j$ at $t=0$, $y_{ij}$ would be equal to $1 - (1-p_0)^2$. The $\frac{r}{r+\lambda_{ij}}\frac{c(d_{ij})}{1-c(d_{ij})}$ term corresponds to interactions that are not maintained due to the initial lack of information. Not surprisingly, as $\lambda_{ij} \rightarrow \infty$ (i.e., as the information is revealed immediately) or as $c(d_{ij}) \rightarrow 0$ (maintaining interaction is not costly), this term goes to zero.

Our first result is a sanity check, which generates an empirical prediction we later verify in Section \ref{section:diagnosing homophily}. It directly follows from Equation \eqref{eqn:yij} and Assumption \ref{assumption:learning}.
\begin{result}[Homophily] \label{result:homophily}
Suppose the pair $(i,j)$ is more similar than another pair $(i,\ell)$. Then, $y_{ij}(d) > y_{i\ell}(d)$ for $d \in \{\underline{d},\overline{d}\}$.
\end{result}
Throughout the remainder of this section, we are interested in the {\it treatment effect of proximity}, defined as:
\begin{align*}
    \gamma_{ij} = y_{ij}(\underline{d}) - y_{ij}(\overline{d}) 
\end{align*}
Here, $\gamma_{ij}$ is simply the causal effect of placing $i$ and $j$ closer on the likelihood of $i$ and $j$ maintaining their interaction. Our second result follows from Equation \eqref{eqn:yij} and Assumption \ref{assumption:cost}.
\begin{result}[Positive Treatment Effects] \label{result:proximity}
For any $i$ and $j$, $\gamma_{ij} > 0$.
\end{result}
Our final result is the main hypothesis we explore empirically. It follows from Assumption \ref{assumption:cost} and \ref{assumption:learning}, and Equation \eqref{eqn:yij}.
\begin{result}[Heterogeneous Treatment Effects] \label{result:heterogeneouseffects}
Suppose the pair $(i,j)$ is more similar than another pair $(i,\ell)$. Then, $\gamma_{ij} < \gamma_{i\ell}$.
\end{result}
Result \ref{result:heterogeneouseffects} implies that, controlling for other dimensions of characteristics, the interaction of proximity and similarity has a negative effect on the likelihood of maintaining interactions. In other words, the causal effect of proximity is more pronounced on dissimilar pairs. Intuitively, this is driven by the substitutability of {\it proximity} and {\it similarity} for fostering friendships. For a given distance, similar pairs explore more and learn about the value of interaction better. Since they already have better information, proximity leaves less room for learning for such pairs. The easiest way to see this is considering the extreme case where $\lambda_{ij} = \lambda_{ji} = \infty$ for similar pairs (i.e., they immediately learn about their values of interaction). In this case, a lower value of distance would have no effect on the likelihood of friendship for similar pairs, because all the learning takes place regardless of $c(d_{ij})$.

We conclude this section by a brief discussion on the normative implications of these results. Consider a school principal who is assigning students $j$ and $\ell$ to two possible locations. Each location can accommodate only one student, one of the locations is at distance $\underline{d}$ from $i$, whereas the other one is at distance $\overline{d}$ from $i$. Suppose the pair $(i,j)$ is more similar than $(i,\ell)$, and the objective of principal is to foster the total number of interactions maintained. Result \ref{result:heterogeneouseffects} means that the optimal placement assigns $\ell$ to the location closer to $i$. This argument can easily be replicated to multiple students, and the main insight would remain: a designer who wants to foster interactions should place dissimilar pairs closer.

\subsection{Alternative Models of Homophily} \label{section:preference-based-homophily}

In our benchmark model, the ex ante expected payoff from an interaction is $p_0$, independent of the pair's similarity. Result \ref{result:homophily}, therefore, is purely driven by different rates of learning. We call this {\it learning-driven homophily}. One way to have Result \ref{result:homophily} without heterogeneous learning is to having heterogeneous payoffs, as in \cite{curr_jackson_pin_ecta} and \cite{mele_ecta}. To this end, we consider an extension of the model where $v_{ij}$ is also heterogeneous and depends on the pair's characteristics. 

Suppose that $v_{ij} \in \{0,\bar{v}_{ij}\}$. At the beginning of encounter, $i$ observes $\bar{v}_{ij}$ and $j$ observes $\bar{v}_{ji} = \bar{v}_{ij}$. Here, $\bar{v}_{ij}$ the {\it maximum} flow payoff $i$ could receive from maintaining interactions with $j$. Nevertheless, $i$ still does not know the realization of $v_{ij}$ and has a prior $\Pr(v_{ij} = \bar{v}_{ij}) = p_0$. To learn the realization of $v_{ij}$, $i$ needs to maintain interactions with with $j$ as defined in the Section \ref{section:process}. Similarly, $v_{ji} \in \{0,\bar{v}_{ji}\}$ is independently drawn with a prior $\Pr(v_{ji} = \bar{v}_{ji}) = p_0$.

To convey our message in simplest terms, let $\bar{v}_{ij}$ have two possible values. Suppose $\bar{v}_{ij} \in \{v_L, v_H\}$ with $v_L < c(\underline{d}) < c(\overline{d}) < v_H = 1$. Here, some friendships have a maximum flow payoff of $v_L$, which is low enough so that they are not worth exploring. The remaining friendships have a maximum flow payoff of $v_H$, in which case the pair explores the value of friendship. We represent $\bar{v}_{ij}$ as a random variable drawn according to the distribution $Pr(\bar{v}_{ij} = v_H) = \mu_{ij}$. To capture heterogeneous payoffs, we impose the following structure on the payoffs.
\begin{assumption} \label{assumption:preference}
Suppose the pair $(i,j)$ is more similar than another pair $(i,\ell)$. Then, $(i,j)$ has a higher maximum payoff on average than $(i,\ell)$, i.e., $\mu_{ij} > \mu_{i\ell}$.
\end{assumption}
Following the same steps as above shows that in this formulation,
\begin{align*}
    y_{ij}(d_{ij}) & = \mu_{ij} \cdot \left( 1 - (1-p_0)^2 \left(1 + \frac{r}{r+\lambda_{ij}}\frac{c(d_{ij})}{1-c(d_{ij})}  \right)^2 \right)
\end{align*}
Under Assumptions \ref{assumption:cost}-\ref{assumption:preference}, Result \ref{result:homophily} still holds: friendship patterns exhibit homophily. In the extended model, there are two factors contributing to homophily. First, heterogeneity in $\bar{v}_{ij}$ generates variation in the {\it extensive margin}: some interactions are not explored at all, and, due to Assumption \ref{assumption:preference}, such interactions are more likely to be amongst dissimilar students. In contrast, heterogeneity in $\lambda_{ij}$ generates homophily due to variation in the {\it intensive margin}: out of explored interactions, similar students explore them longer and learn their payoffs more precisely. The former is the {\it preference-driven homophily}, whereas the latter is {\it learning-driven homophily}.

Regarding the analogous version of Result \ref{result:heterogeneouseffects}, the relative strength of extensive versus intensive margin determines the direction of the heterogeneous effect of proximity. If the variation in extensive margin (i.e., preference-driven homophily) is relatively weaker, Result \ref{result:heterogeneouseffects} still holds. However, if the extensive margin is strong enough, the opposite of Result \ref{result:heterogeneouseffects} holds. Our empirical results in the next section suggest that preference-driven homophily is relatively weaker, and learning-driven homophily is the dominant.

\section{Experimental Setting}
\label{sec: setting}

We now turn to the empirical analysis, where we leverage an experiment that randomly varies the physical proximity between students to test the theoretical predictions in Section \ref{sec: model}. The setting of the experiment is the COAR (\textit{Colegios de Alto Rendimiento}) Network in Peru. The COAR Network is a set of public selective boarding schools designed for the most talented low-income students in the country. They operate for the last three years of secondary school. 

The COAR Network comprises twenty-five schools and enrolls approximately three thousand students every year. Students ages typically range from 14-15 at school entry to 17-18 at graduation. As these are boarding schools, students spend more time in the school relative to day schools. The schools operate Monday through Friday from approximately 7:30 a.m. to 3:45 p.m. and Saturday from 7:30 a.m. to 12:45 p.m. Outside school hours, students can study, play with classmates, and do homework in their dormitories. Students can visit their families on weekends as long as a family member can pick them up from school. Because many students come from a different region than the school, some stay at the school on weekends, which offers new opportunities to interact with peers.

Applicants are eligible for admission to COAR if they are ranked in the top ten of their public school cohort in the previous academic year. The admission decisions are within the region of origin and depend on a composite score of three tests: a math and reading test, a one-on-one interview, and the observation of students' interactions with other applicants.

The three cohorts in our study entered the COAR Network between 2015 and 2017. The 2017 allocation to dorms was determined by randomly assigning students' names to a list, and the position on the list predicts the physical proximity between the students. 

In December 2016, we collected a baseline survey of social networks. The survey included questions on preferred roommates, friendships, study partnerships, and other social activities (such as playing sports or games). After the experiment, we collected two follow-up social network surveys with the same questions (August and December 2017) that we use to test the impact of physical proximity on social connections. 

\subsection{Descriptive Statistics}

Although schools in the COAR Network are restricted to students enrolled at public schools before admission, there is substantial variation in their demographic characteristics. Social workers perform a survey upon the entrance of students, classifying them as poor or non-poor. We restrict our sample to the 19 schools for which this information is available. Table \ref{tab: summary statistics} presents the summary statistics of students in our sample. Approximately 43\% of the students enrolled are male, around 25\% come from rural households, and 41\% are classified as poor. 

We also characterize students' academic achievement and social centrality at baseline. First, we classify students into lower- or higher-achieving types. In particular, within the same school, grade, and gender, students with an admission score above (below) the cell-specific median are classified as higher-achieving (lower-achieving). Second, we classify students as either less socially central or more socially central, based on the  \textit{eigenvector centrality}\footnote{Eigenvector centrality measures a student's influence within their social network. High values indicate that a student is connected to many other individuals who also have high values.} of the social network at the baseline survey. Students with a centrality above (below) the cell-specific median are classified as more (less) socially central. Because the social network surveys were not available for students who enrolled in 2017, these students are not classified as less or more socially central. Hence, the analysis of social centrality focuses on the 2015 and 2016 cohorts. Table \ref{tab: summary statistics} presents the distribution of student types. By construction, half of the students are higher- or lower-achieving, and half are more or less socially central.

Table \ref{tab: summary statistics} also presents the summary statistics of the number of connections students have after the experiment. As in the theoretical section (Section \ref{sec: model}), we consider two students to be connected if, in either of the two follow-up surveys, either of them names the other as a friend, a study partner, or someone with whom they play sports or engage in other social activities. On average, a student has 10.5 friends, 6.6 study partners, and 8.2 social activities partners. 
These account for a total of 14.1 connections. Columns 2 and 3 show that poor and non-poor students have a similar number of links. Columns 4 and 5 show that lower-achieving students have approximately 0.5 fewer connections than higher-achieving students. Although not surprising, the most striking comparison is in columns 6 and 7: less-central students have, on average, 3.5 fewer connections than the more-central students. 

\subsection{Diagnosing Homophily} \label{section:diagnosing homophily}

In light of Result \ref{result:homophily}, we explore whether social networks after the intervention exhibit homophily by poverty, academic achievement, and centrality. We estimate the following equation:
\begin{equation}
    y_{i} = \alpha + \beta_{1} \text{poor}_{i} + \beta_{2} \text{lower-achieving}_{i} + \beta_{3} \text{less-central}_{i} + \delta y_{{i}_{b}}+\sum_{c \in \mathcal{C}} \gamma_{c} d_{ic}  + \varepsilon_{i},
    \label{eq: empirical homophily}
\end{equation} 

\noindent where $y_{i}$ is the number of connections of individual $i$. Our outcomes include the total number of connections a student has after the intervention, as well as the number of connections with students of different types. The variables $\text{poor}_{i}$, $\text{lower-achieving}_{i}$, and $\text{less-central}_{i}$ are dummy variables that take the value of one when individual $i$ is poor, lower-achieving, and less central, respectively. We control for the number of connections at baseline $ y_{{i}_{b}}$, and include cell (school-by-grade-by-gender) fixed effects captured by the dummy variables $d_{ic}$. Finally, $\varepsilon_{i}$ is an error term.

Table \ref{tab: homophily} reports the estimates of Equation \eqref{eq: empirical homophily}. The first column reports the estimates on the total number of connections. Poor students have on average 0.55 fewer  connections (p-value $<0.01$) than non-poor students. While lower-achieving students have on average 0.22 fewer links, this difference is statistically indistinguishable from zero. Less central students have 1.14 fewer links than more central students (p-value $<0.01$).

Students have more connections with those similar to them. Columns 2 and 3 show that while poor and non-poor students have a similar number of connections with other poor students, there is a difference of 0.66 links (p-value$<0.001$) with non-poor students. Columns 4 and 5 show that compared to higher-achieving students, lower-achieving students have around 1.06 more links with other lower-achieving students and around 1.27 fewer links with higher-achieving students. Finally, the results in columns 6 and 7 show that compared to the more-central students, less-central students form around 0.23 more links (p-value$=0.035$) with other less-central peers and 1.07 fewer links (p-value $< 0.001$) with more-central peers. In summary, social networks exhibit homophily in academic achievement, poverty, and social centrality. 

\section{The Role of Proximity}

\subsection{Random Variation in Proximity}

The experiment generates random variation in the physical proximity of two students in the allocation to dormitories. In particular,  students' names were randomly sorted on a list, and the schools followed the list to allocate students to specific beds in the dormitories. The order on the list predicts the physical distance between two students in dormitories. The details of this design are explained in \citet{zarate_jmp} and in Appendix \ref{appendix: allocation}, where we also present evidence that the distance on the list predicts the likelihood that students are neighbors in dormitories. 

We leverage the random variation generated by the list to estimate the impact of proximity on the likelihood that individuals $i$ and $j$ form a social connection. In particular, we estimate the following equation to explore this relationship:
\begin{equation}
y_{ij}=\alpha + \gamma l^{d}_{ij} + \eta l^{d}_{ij} \times first_{ij} + \omega' X_{ij} + \phi_{i} + \phi_{j} + \varepsilon_{ij} ,
\label{eq: proximity}
\end{equation}
\noindent  where $y_{ij}$ is a dummy variable equal to one when students $i$ and $j$ form a social connection. The variable $l^d_{ij}$ is a dummy variable equal to one when student $j$ is in the neighborhood of size $d$ of student $i$ on the list. In other words, whether the difference between the position or the row of students $i$ and $j$ on the list is lower than $d$:
\begin{align*}
    l^{d}_{ij}= 
     \begin{cases}
1     & \text{if } distance(i, j) \leq d,\\
0     & \text{otherwise.}
\end{cases}
\end{align*}

We also include an interaction term between $l^{d}_{ij}$ and a dummy variable for first-year students in Equation \eqref{eq: proximity}. We include this interaction as first-years did not know each other before the intervention, and some schools also use the lists for classroom allocation. The vector $X_{ij}$ includes covariates varying at the link level. These include the combination of gender fixed effects as there is homophily by gender and dorm allocation depends on gender, the combination of type (by academic achievement and social centrality) fixed effects as this was part of the experimental design, and whether $i$ and $j$ had a social connection at baseline. We also control for ego ($\phi_{i}$) and alter ($\phi_{j}$) fixed effects. Finally, $\varepsilon_{ij}$ is an error term. We cluster the standard errors at the network (school-by-grade) level. 

\begin{figure}[h!]
    \caption{Effect of Proximity on Social Interactions}
    \begin{center}
    \scalebox{0.8}{\begin{tabular}{cc}
\multicolumn{2}{c}{Panel A: Effect of Proximity}    \\
\multicolumn{2}{c}{Before the Intervention} \\
\multicolumn{2}{c}{\includegraphics[scale=0.6]{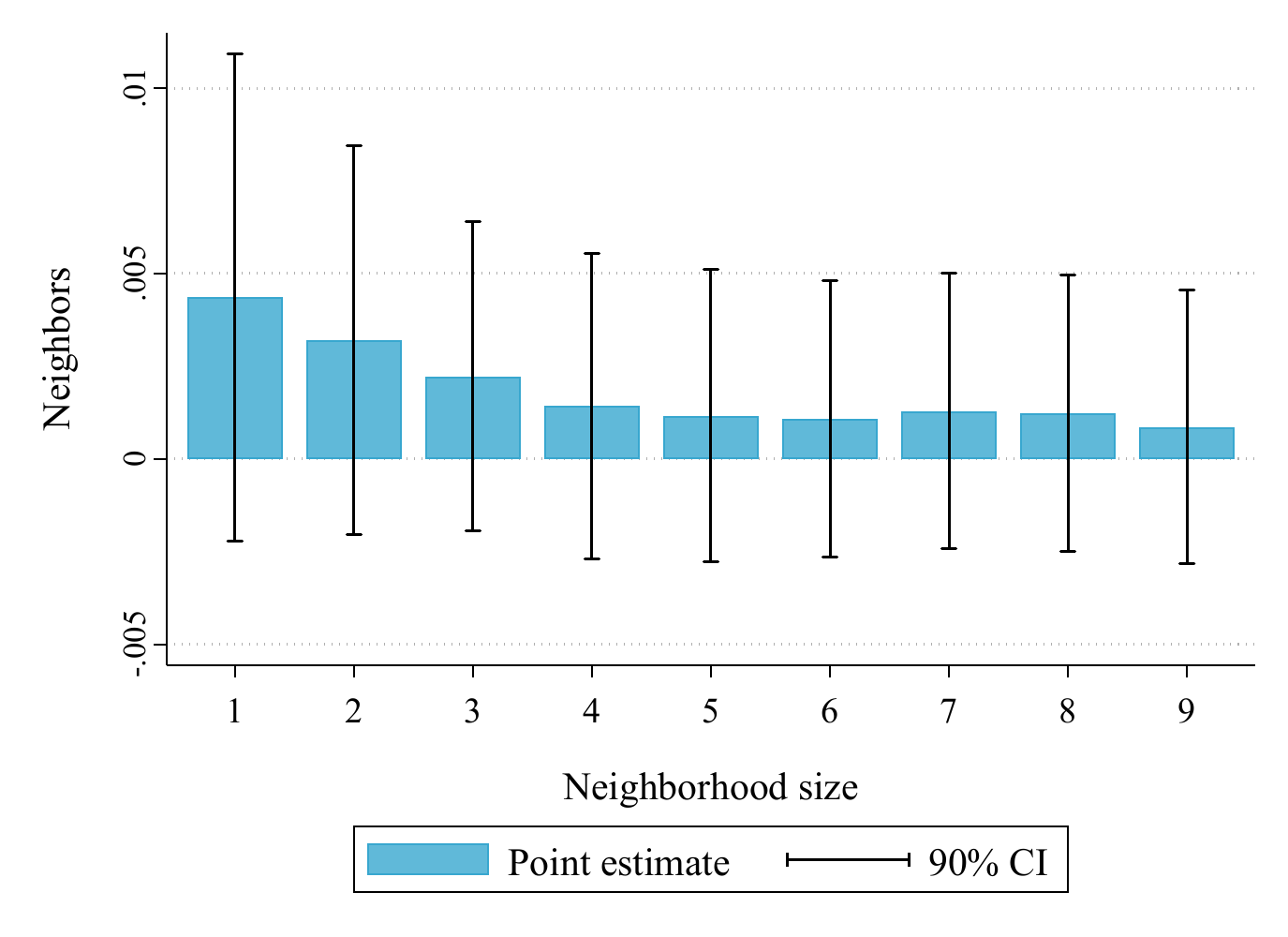}}    
\\
\multicolumn{2}{c}{After the Intervention}  \\
Panel B: Effect of Proximity & Panel C: Interaction with First-year Dummy \\
\includegraphics[scale=0.6]{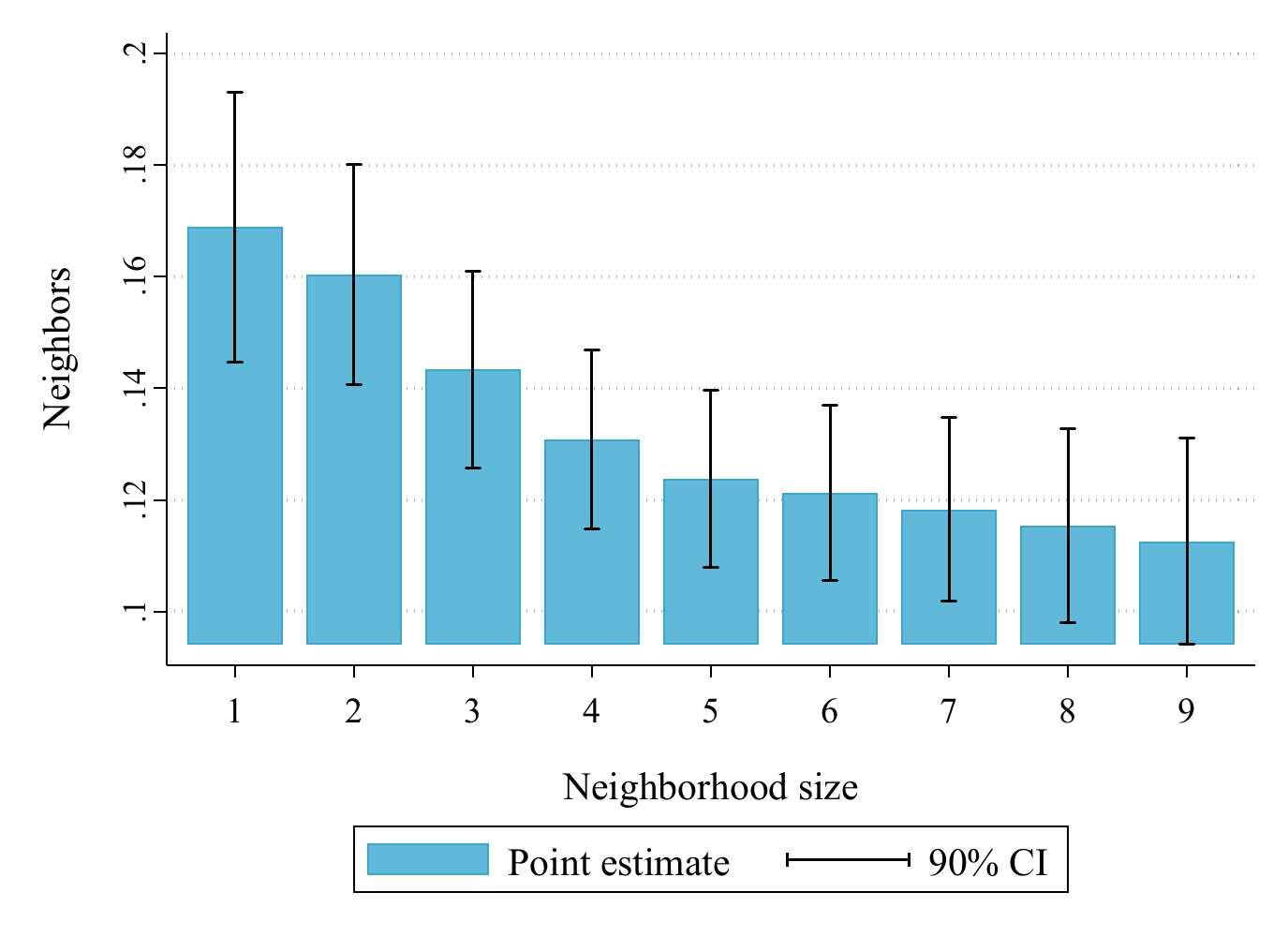} &     \includegraphics[scale=0.6]{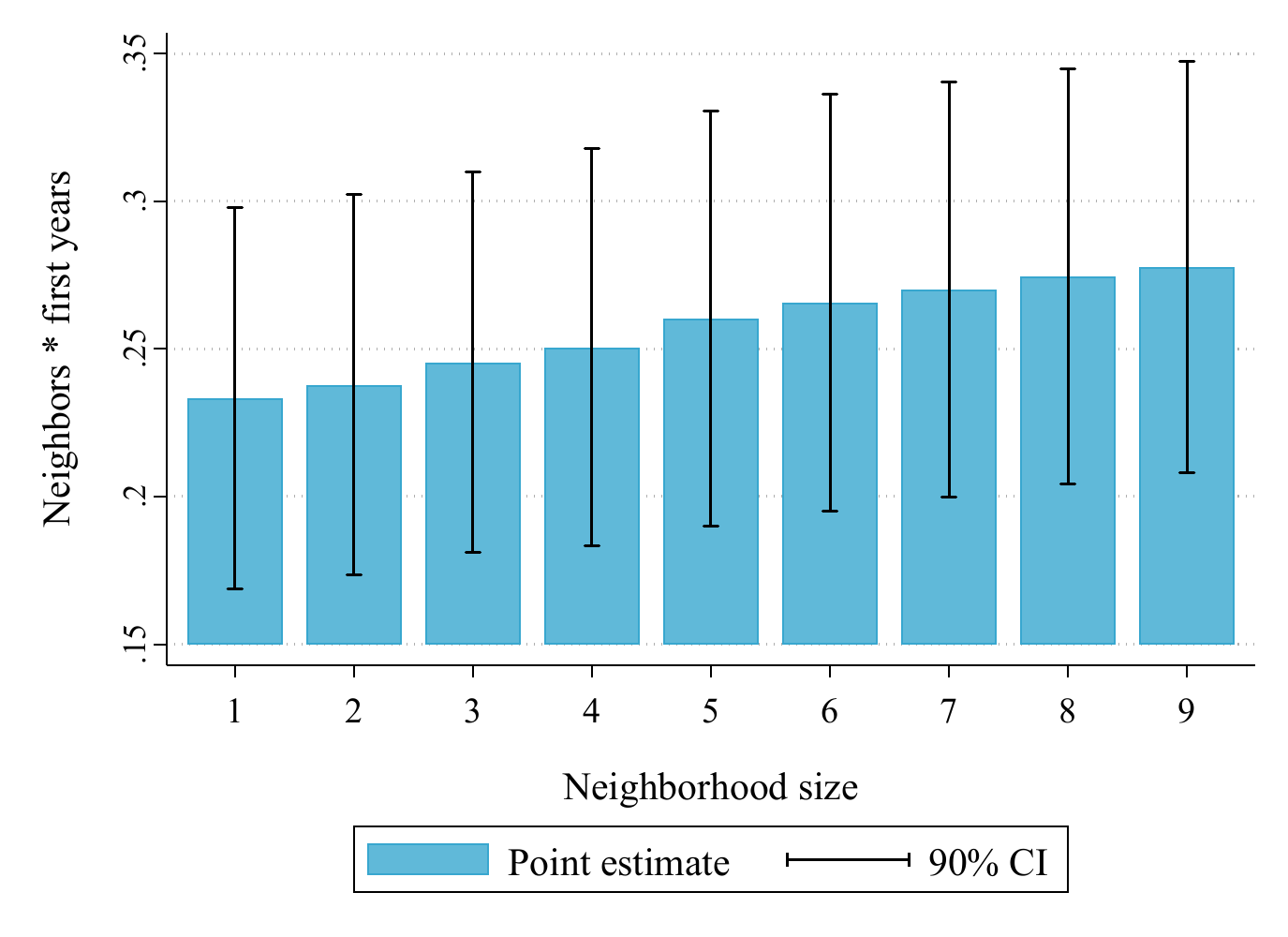} \\
    \end{tabular}}
    \end{center}
    \footnotesize{\textbf{Notes}: This figure presents estimates and 90\% confidence intervals of Equation \eqref{eq: proximity} for neighborhoods of different sizes on the list. Panel A shows the direct effect of proximity on social connections at baseline,  $N=213,799$. Social connections at baseline are not available for first-year students. Panels B and C report estimates on social connections after the intervention aggregating the networks at midline and endline. Panel B reports the direct effect of proximity, and Panel C the additional effect for first-year students, $N=348,079$. Standard errors are clustered at the network (school-by-grade) level. \par}
    \label{fig: proximity}
\end{figure}

Figure \ref{fig: proximity} reports the estimates of Equation \eqref{eq: proximity}. Panel A shows the results for social interactions before the intervention.\footnote{This data at baseline is not available for the first-year students} In general, and as expected from the random variation, proximity does not predict social interactions before the intervention for all the neighborhood sizes. 

Proximity does predict social interactions after the intervention. Panels B and C present these estimates. The results in Panel B show that being adjacent on the list (a distance of 1) increases the likelihood of social interactions by approximately 17 percentage points (p-value$<$0.01). As we increase the neighborhood size, the effect becomes lower, although it is still statistically significant for a neighborhood of size 8. 

The estimates in Panel C show that the effect of proximity is higher for first-year students. For example, when first-year students' names are adjacent on the list, there is an increase of 23 percentage points of them forming a social connection compared to students in other grades. This larger effect is intuitive as first-year students have not created previous social links. They are also more likely to share the same classroom.

\subsection{Heterogeneous Effects of Proximity}

We now explore whether there are heterogeneous effects of proximity by students' characteristics. In particular, we consider whether students from different poverty statuses, achievement levels, and centrality at baseline form more or less social connections with their neighbors than students that share these characteristics. Result \ref{result:heterogeneouseffects} predicts that the effect of proximity is higher for dissimilar students. We estimate the following equation: 
\begin{equation}
\begin{array}{rcl}
y_{ij}&=&\alpha + \beta_{p} D_{p_{ij}} + \beta_{a} D_{a_{ij}} + \beta_{s} D_{s_{ij}} + \\
& & \gamma l^{d}_{ij} + \delta_{p} l^{d}_{ij} \times D_{p_{ij}} + \delta_{a} l^{d}_{ij} \times D_{a_{ij}} + \delta_{s} l^{d}_{ij} \times D_{s_{ij}}+ \\
&&\omega' X_{ij} + \phi_{i} + \phi_{j} + \varepsilon_{ij} ,
\end{array}
\label{eq: heterogeneity}
\end{equation}

\noindent where $D_{p_{ij}}$, $D_{a_{ij}}$, and $D_{s_{ij}}$ are dummy variables equal to one when students $i$ and $j$ have a different poverty status, achievement level, and centrality at baseline, respectively. The remaining variables and parameters are as in Equation \eqref{eq: proximity}. The parameters $\bm{\beta} \in \{\beta_p, \beta_a, \beta_s\}$ capture whether social networks exhibit homophily. Specifically, when $\beta<0$ students of different types are less likely to form social interactions.

Our parameters of interest are $\bm{\delta} \in \{\delta_{p}, \delta_{a}, \delta_{s}\}$, which capture whether the effect of proximity is weaker or stronger when students are of a different type. Based on the results in Section \ref{sec: model}, learning-based homophily predicts that $\delta>0$, whereas preference-based homophily predicts $\delta<0$.

\begin{figure}[h!]
    \caption{Heterogeneous Effects of Proximity on Social Interactions}
    \begin{center}
    \scalebox{0.8}{\begin{tabular}{c}
    Panel A: Effect of Proximity  \\
         \includegraphics[scale=0.6]{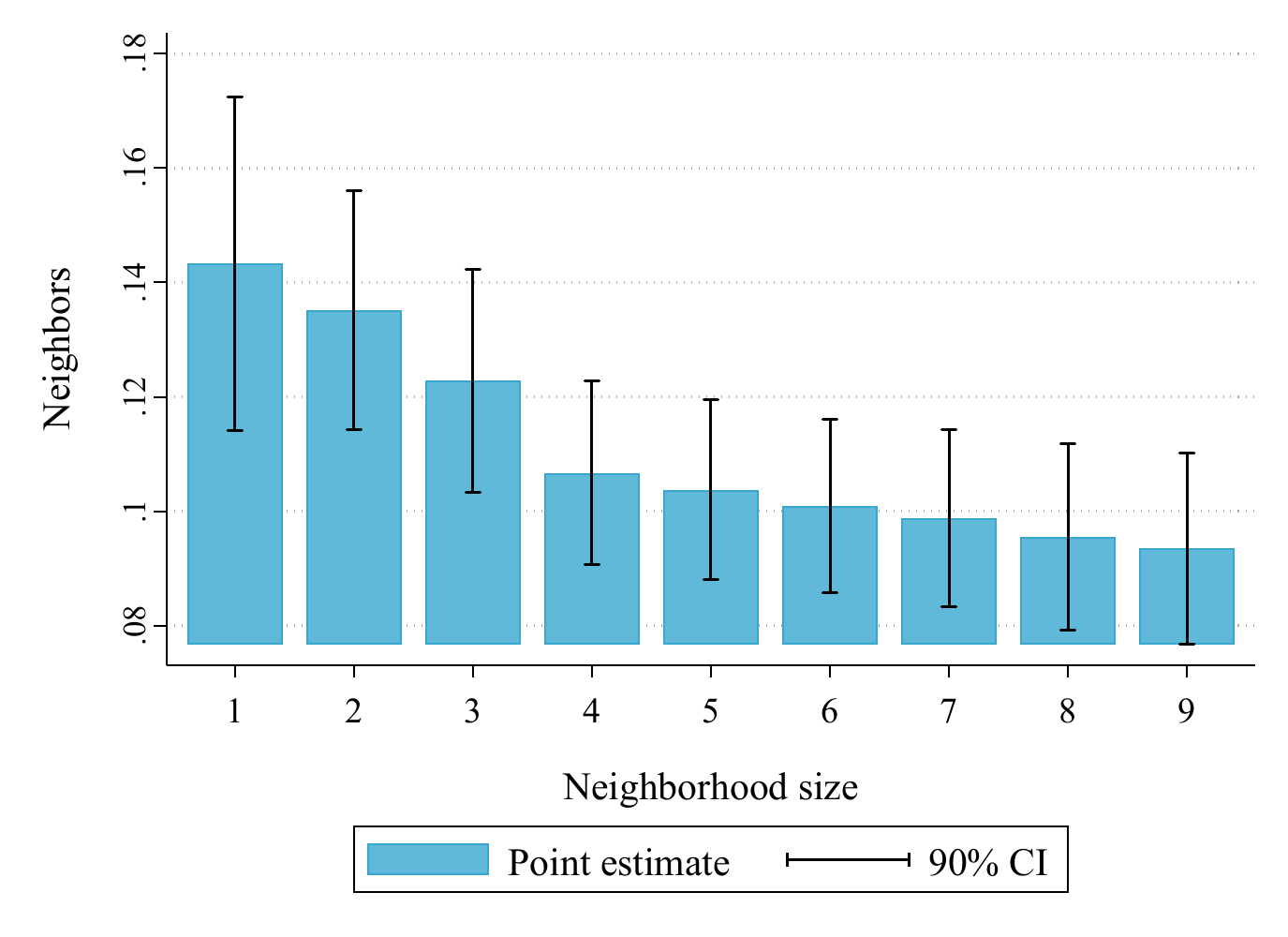} 
    \end{tabular}}
    \scalebox{0.8}{\begin{tabular}{cc}
    Panel B: Interaction with      &  Panel C: Interaction with    \\
         First-year Dummy     &   Different Poverty Statuses Dummy   \\
         \includegraphics[scale=0.6]{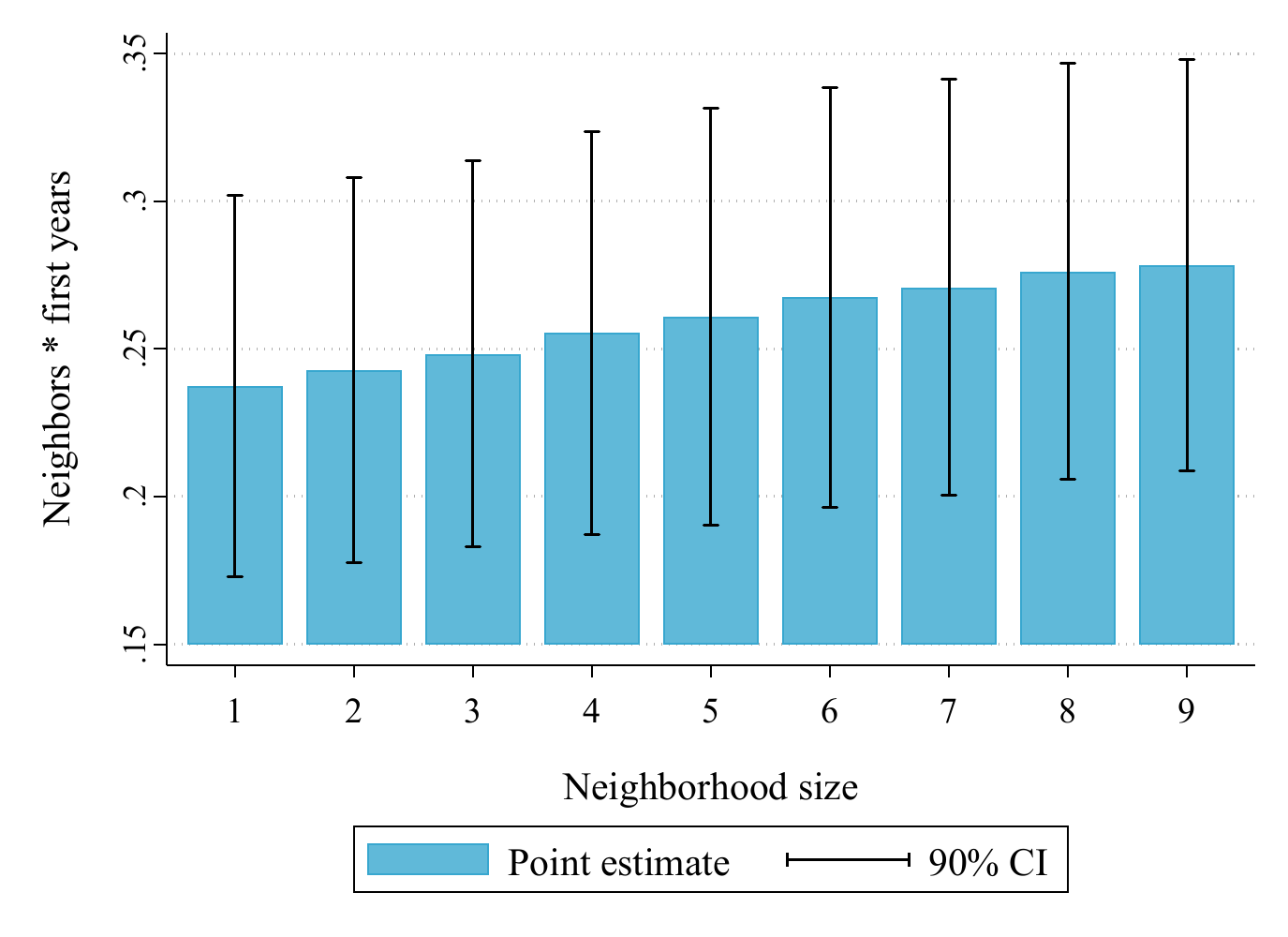} &
         \includegraphics[scale=0.6]{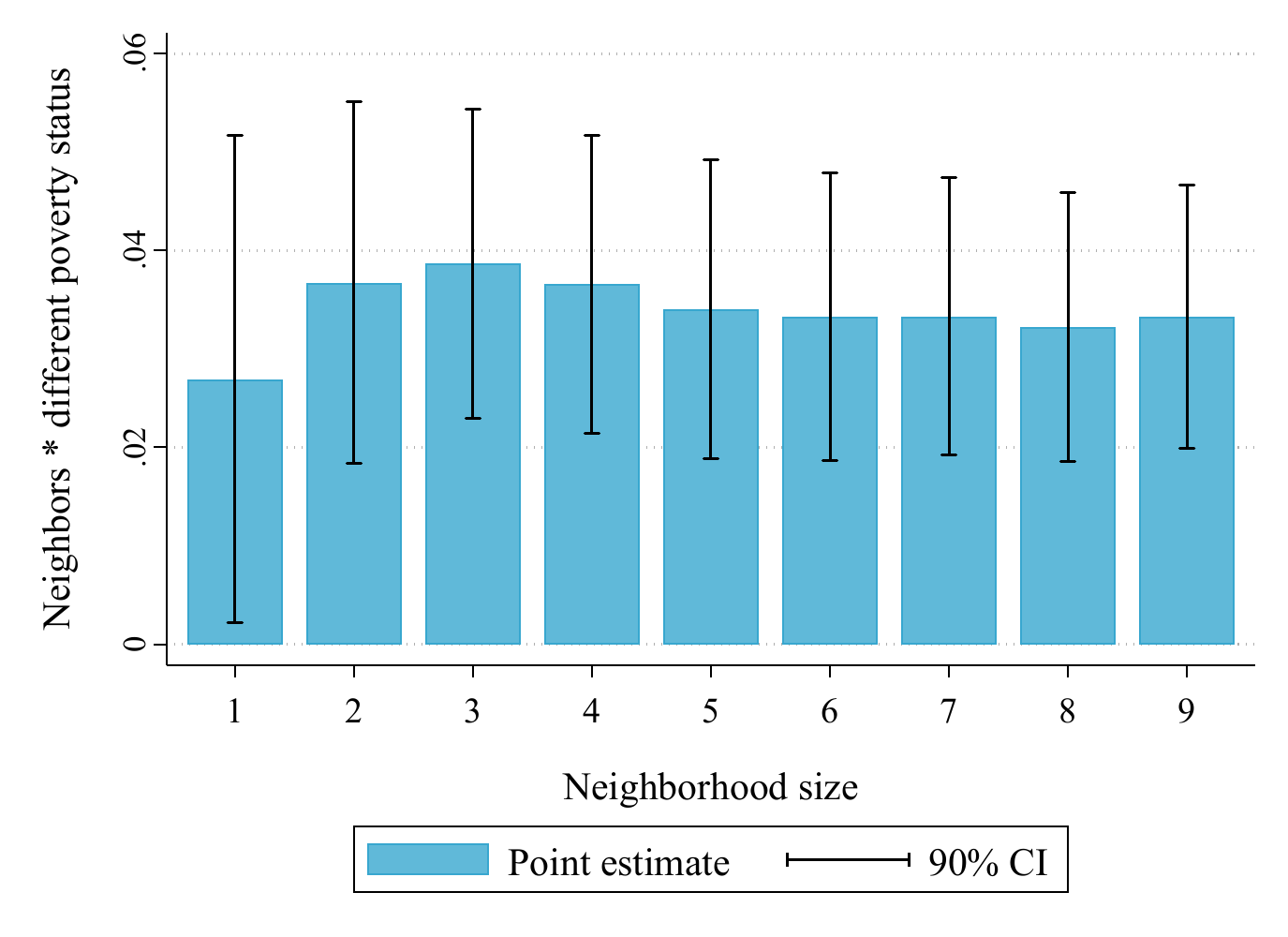} \\
         Panel D: Interaction with   &  Panel E: Interaction with  \\
         Different Achievement Levels Dummy &   Different Social Centrality Dummy \\
         \includegraphics[scale=0.6]{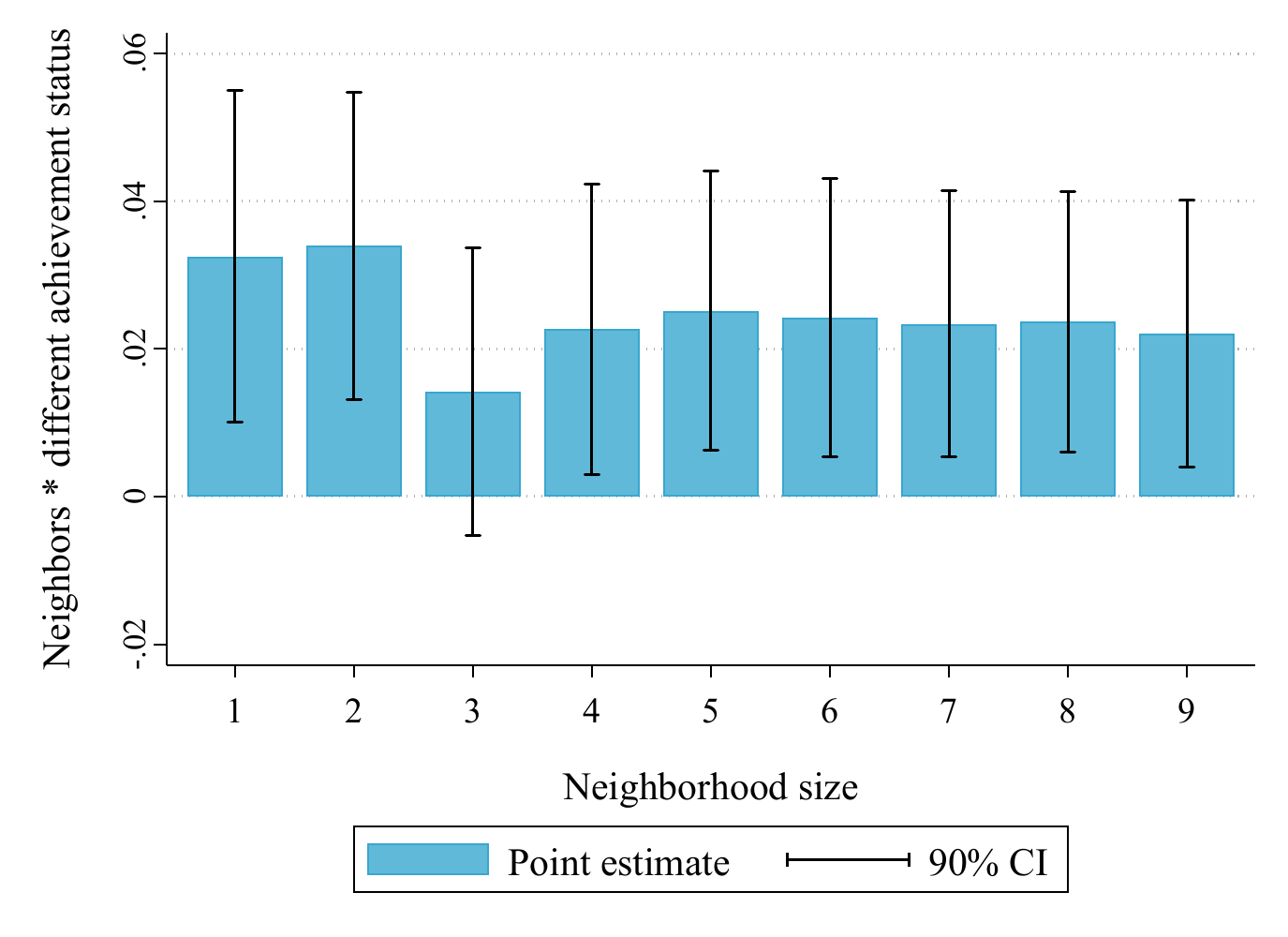} &
         \includegraphics[scale=0.6]{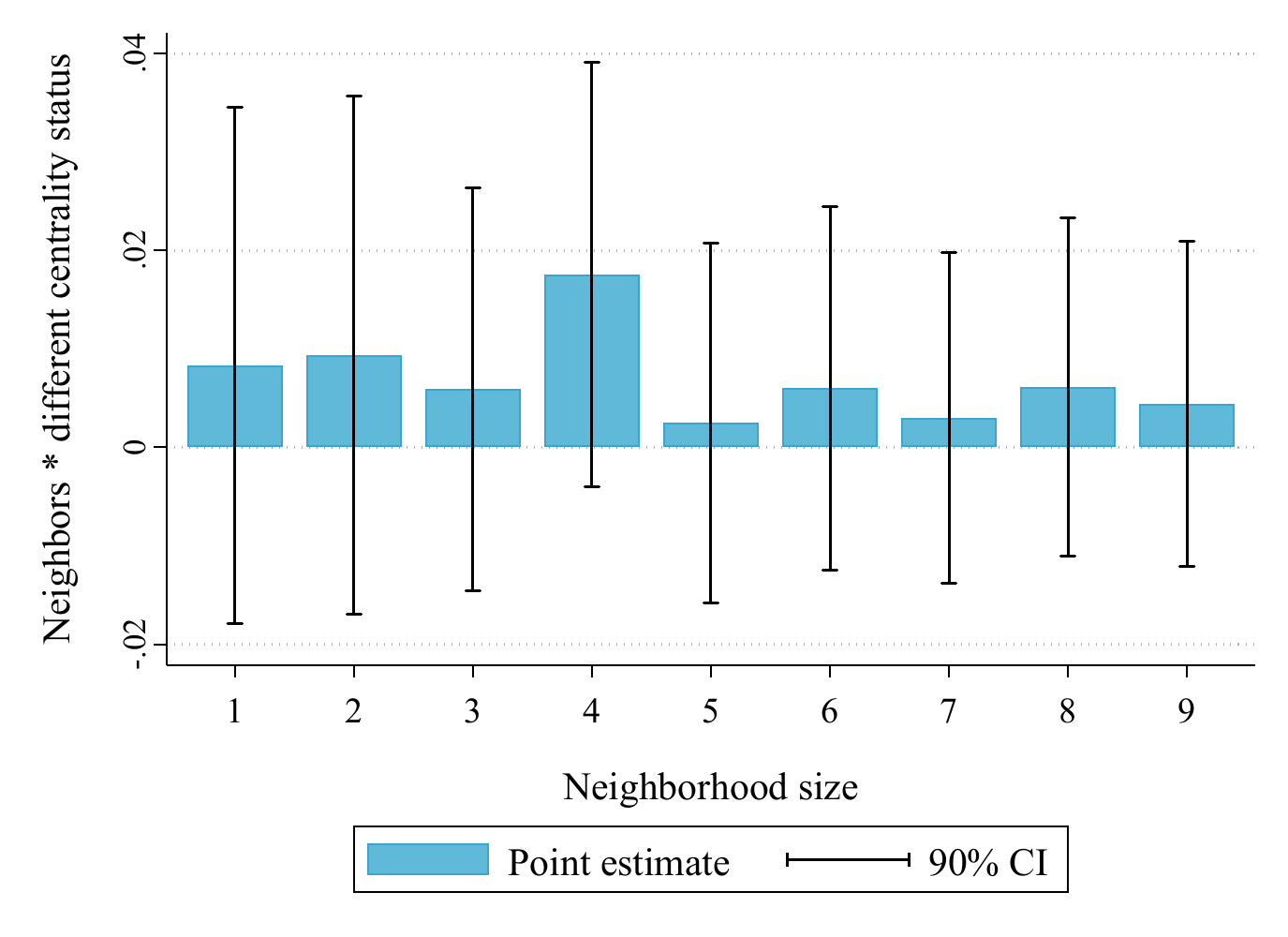} 
    \end{tabular}}
    \end{center}
    \footnotesize{\textbf{Notes}: This figure reports estimates and 90\% confidence intervals of Equation \eqref{eq: heterogeneity} for neighborhoods of different sizes on the list. The dependent variable is forming a social connection after the intervention aggregating social networks at midline and endline, $N=348,079$. Panel A reports the direct effect of proximity (being at a distance less or equal to $d$ on the list). Panel B presents the additional effect of proximity for first-year students. Panels C, D, and E report the interaction with whether students have a different poverty status, academic achievement, and social centrality, respectively. Standard errors are clustered at the network (school-by-grade) level. \par}
    \label{fig: heterogeneity}
\end{figure}

Figure \ref{fig: heterogeneity} reports the estimates of Equation \eqref{eq: heterogeneity}. Overall, the estimates are consistent with a learning-based model of homophily. The impact of proximity is higher for students that differ in their poverty status and achievement level. Panels C-E display the estimates of the vector $\bm{\delta}$ for neighborhoods of different sizes: the heterogeneous impact of proximity for students with different poverty statuses, achievement levels, and social centrality levels. The figures in Panels C and D show that we can reject at conventional significance levels that $\delta_{p}$ and $\delta_{a}$ are equal to zero for neighborhood sizes from 1 to 9. Students are more likely to befriend those with a different poverty status (3.4 percentage points, on average) and achievement levels (2.4 percentage points, on average) when physically close in dormitories. By contrast, we cannot reject a homogeneous impact of proximity for students that share or differ in their social centrality at baseline. The poverty and achievement level results are consistent with the learning-driven homophily results derived in Section \ref{sec: model}.  

\section{Conclusion}

This paper introduces a model of friendship formation that can explain patterns of homophily in social networks. The model predicts that physical proximity and similarity are substitutes when students decide how to form social connections. Hence, the effect of proximity would be larger for dissimilar students.

We corroborate the model by leveraging an experiment at selective boarding schools in Peru that generates random variation on the physical proximity between students. We interpret this variation as changes in the cost of forming social connections. The empirical results show that social networks exhibit homophily in poverty, academic achievement, and baseline social centrality. The effect of proximity on social interactions is larger for students with different poverty statuses and academic achievement levels. 

Considering the effects of within-school policy variation is vital to understand the consequences of school integration policies. Even though social networks in the COAR Network exhibit homophily, policies that foster diversity by mixing students of different types in dormitories can enhance the structure of more diverse social networks.

\newpage

\linespread{1}

\begin{landscape}

\begin{table}
\caption{Summary Statistics}
    \begin{center}
    \scalebox{0.8}{\begin{tabular}{lcccccccccc}
        \hline
    \hline
    Variable & All students & & \multicolumn{2}{c}{By poverty} & & \multicolumn{2}{c}{By Academic Achievement} & & 
    \multicolumn{2}{c}{By Social Centrality} \\
    \cline{4-5} \cline{7-8} \cline{10-11}
    & & & Poor students & Non-poor students & & Lower-achieving & Higher-achieving & & Less central & More central \\
    & (1) & & (2) & (3) & & (4) & (5) & & (6) & (7) \\
    \hline
\textit{Demographics} \\
Male (\%)&0.43&&0.44&0.43&&0.43&0.43&&0.41&0.40\\
Rural (\%)&0.25&&0.43&0.14&&0.29&0.21&&0.29&0.23\\
Poor (\%)&0.41&&1.00&0.00&&0.46&0.36&&0.47&0.39\\
\\
\textit{Network variables} \\
   Connections&14.15&&14.19&14.95&&13.93&14.37&&11.51&14.97\\
&6.85&&6.23&6.59&&6.90&6.78&&5.55&5.99\\
Connections with poor students&5.54&&6.91&4.89&&5.63&5.45&&4.86&6.08\\
&(4.11)&&(4.42)&(3.54)&&(4.17)&(4.05)&&(3.81)&(4.48)\\
Connections with lower-achieving students&7.03&&7.26&7.29&&7.54&6.53&&5.76&7.34\\
&(4.25)&&(3.99)&(4.23)&&(4.57)&(3.84)&&(3.31)&(3.57)\\
Connections with less central students&9.58&&9.43&10.22&&9.50&9.66&&5.76&5.96\\
&(7.23)&&(6.73)&(7.39)&&(7.21)&(7.25)&&(3.24)&(3.34)\\

\\
N&5,467&&2,148&3,063&&2,731&2,736&&1,696&1,687\\
    \hline
    \hline
    \end{tabular}}
    \end{center}
    \footnotesize{\textbf{Notes}: This table reports summary statistics for demographic characteristics and the total number of connections students have with their peers. The demographic characteristics  include whether the student is a male, whether they come from a rural household, or from a household classified as poor. Column 1 reports the statistics for all students enrolled at the COAR Network in 2017. Columns 2 and 3 classify students by poverty status, columns 4 and 5 by academic achievement, and columns 6 and 7 by baseline social centrality. The standard deviations are in parenthesis.   \par}

    \label{tab: summary statistics}
\end{table}

\clearpage

\begin{table}
\caption{Estimates of Students' Characteristics on Number and Type of Connections }
    \begin{center}
    \scalebox{0.82}{\begin{tabular}{lcccccccccc}
        \hline
    \hline
    & \multicolumn{10}{c}{Number of connections with:} \\
    & All students & & Poor students & Non-poor students & & Lower-achieving & Higher-achieving & & Less central & More central \\
    & (1) & & (2) & (3) & & (4) & (5) & & (6) & (7) \\
    \hline
    \input{tables/table2_tex.txt}
    \hline
    \hline
    \end{tabular}}
    \end{center}
    \footnotesize{\textbf{Notes}: This table reports estimates of the impact of students' characteristics on the number and type of connections. Specifically, the estimates show how poor, lower-achieving, and less socially central students have different connections and links with specific types of students. Standard errors are clustered at the peer-group-type-by-student-type level; *** p-value$<$0.01, ** p-value$<$0.05, * p-value$<$0.1.    \par}

    \label{tab: homophily}
\end{table}

\end{landscape}

\clearpage

\clearpage

\bibliographystyle{econ}
\bibliography{references.bib}

\clearpage
\appendix

\setcounter{table}{0}
\setcounter{figure}{0}
\renewcommand\thefigure{\thesection.\arabic{figure}} 
\renewcommand\thetable{\thesection.\arabic{table}} 

{\Large\bf\noindent Appendix}

\section{Experimental Design}
\label{appendix: allocation}

The experiment described in detail in \cite{zarate_jmp} estimates peer effects on academic achievement and social centrality. The design classifies students into types using the median of baseline achievement and social centrality. Conditional on the student's type, they are randomized to the type of peer. For the design to be flexible across dorms of different sizes and structures, students' names are organized on a list based on the randomization to the type of peer. School administrators follow the lists to assign students to specific beds and dorms in the school. This appendix provides a general overview of the experimental design and how it generates random variation in students' physical distance (proximity). 

The experimental design guarantees strong variation in the peer attributes of interest by proceeding in three steps. 
\begin{enumerate}[1.]
\item First, the researcher classifies students into peer types determined by the quantiles in the distribution of the peer attribute of interest. In the simplest case, the classification is determined by the median, with only two peer types\footnote{Since there is a trade-off between the number of treatments arms and statistical power, I implement the simplest design with just two types of peers using the median.}: high and low-type students. 
\item In the second step, conditional on a student's type, each student is assigned to a peer type; in the case of this experiment, either to a high-type-peer treatment (matched with high-type students) or to the control group (matched with low-type students). These two treatment arms are equivalent to assigning students to combinations of a student's and a peer type. These combinations guarantee the treatment's predictive power on peer attributes (a strong first stage) as they vary the proportion of each type: 0\%, 50\%, or 100\%.
\item Third, student names are organized on a list that are used for the students' allocation to groups. In the case of our experiment, to dormitories of different sizes. Lists are determined by combinations of student-peer types and are adaptable to dorms of various sizes.
\end{enumerate}

In the simplest case, with two types of students, high- and low-type students are assigned to high- and low-type peers. If we look at a student and any of her peers in the research design, there would only be three combinations of a student's own type and the peer type.
\begin{enumerate}[a)]
\item Combination A: composed of high-type students assigned to high-type peers.
\item Combination B: a mixed combination where half of the members are high-type students assigned to low-type peers and the other half are low-type students assigned to high-type peers.
\item Combination C: composed of low-type students assigned to low-type peers.
\end{enumerate}

The following matrix illustrates the composition of these combinations, which are a function of a student’s type and her assigned peer type:
\begin{center}
\begin{tabular}{llc|c|}
\cellcolor{white} \multirow{10}{*}{\STAB{\rotatebox[origin=c]{90}{Student Type}}} & \cellcolor{white}   & \multicolumn{2}{c}{Peer Type}  \\
 \cline{2-4}
& \multicolumn{1}{|c|}{$\;$}  & \multicolumn{1}{|c|}{High} &  \multicolumn{1}{|c|}{Low} \\
 \cline{2-4}
& \multicolumn{1}{|c|}{High} & \cellcolor{g1!50}
\begin{tabular}{c}
\underline{Combination A}  \\
Proportion High=100\% \\
Proportion Low=0\%
\end{tabular}
  & \cellcolor{colortl1!70!white}
\begin{tabular}{c}
\underline{Combination B}  \\
Proportion High=50\% \\
Proportion Low=50\%
\end{tabular}
  \\
 \cline{2-4}
&  \multicolumn{1}{|c|}{Low} & \cellcolor{colortl1!70!white}
 \begin{tabular}{c}
\underline{ Combination B}  \\
Proportion High=50\% \\
Proportion Low=50\%
\end{tabular}
& \cellcolor{g10!50}
\begin{tabular}{c}
\underline{Combination C}  \\
Proportion High=0\% \\
Proportion Low=100\%
\end{tabular}
\\
\cline{2-4}
\end{tabular}
\end{center}

Each row in this matrix represents a type of student, and each column is the assigned peer type. The diagonal of the matrix shows all combinations composed of a single student type. Off-diagonal elements of this matrix are symmetrical, as students are matched to peers of the opposite type in Combination B.\footnote{Notice that for all three combinations to have the same size, two-thirds of students are assigned to peers of their same type, and one-third to the mixed combination.} The size of each of these combinations is determined by the sample size of randomization strata. For example, if the total sample is thirty students, fifteen students are high-type, fifteen are low-type, and each of these combinations would have ten students. Combination A would have ten high-type students, Combination B, five high- and five low-type students, and Combination C, ten low-type students. 

More relevant to this paper, this experimental design is flexible enough to adapt to groups of various sizes. In particular, participants' names are sorted on a list based on their student-peer type combination, and schools use the list to assign students to specific groups. In our case, dorms or beds in large dormitories. Under this design, the position on the list predicts the final assignment to groups and the physical proximity between two students in a dorm. For example, students whose names are adjacent on the list are more likely to be roommates or in neighboring beds.

Each student's position on the list is random and determined by the assignment to the treatment as follows. First, student-peer-type combinations are randomly ordered on the list. Second, the students' order in the list is also randomized with one condition: the list alternates the two student types in the mixed combination (Combination B). This rotation guarantees that the closest neighbors' type on the list coincides with the student's treatment arm. For example, for a student assigned to type-$H$ peers, the two adjacent names on the list (the one before and the one after) would be of type $H$.  

Let's consider an example with 12 students (6 low and 6 high types) who get assigned either to the treatment (high-type peers) or the control (low-type peers). First, the student-peer-types combinations are randomly ordered on the list, and one of six potential orders is selected.\footnote{The six potential orders are: (i) A-B-C, (ii)  A-C-B, (iii) B-A-C, (iv) B-C-A, (v) C-A-B, (vi)  C-B-A.} In this example, I assume that the selected random order is A-C-B. Within each combination, students are randomly ordered while adhering to the condition that students in the mixed group alternate. The following list illustrates this example, with the letters $H$ and $L$ representing the student's type and the blue font identifying students assigned to the treatment (high-type peers).
$$\underbrace{\textcolor{blue}{H}-\textcolor{blue}{H}-\textcolor{blue}{H}-\textcolor{blue}{H}}_\text{Group A}-\underbrace{L-L-L-L}_\text{Group C}-\underbrace{H-\textcolor{blue}{L}-H-\textcolor{blue}{L}}_\text{Group B}$$

This illustrative list represents how the experimental design is adaptable in allocating students to dorms of various sizes. For example, if students were assigned to six dorms of two students each, the assignment would look as follows:
$$\underbrace{\textcolor{blue}{H}-\textcolor{blue}{H}}_\text{\color{black}{Dorm 1}}-\underbrace{\textcolor{blue}{H}-\textcolor{blue}{H}}_\text{\color{black}{Dorm 2}}
-\underbrace{L-L}_\text{\color{black}{Dorm 3}}-\underbrace{L-L}_\text{\color{black}{Dorm 4}}-
\underbrace{H-\textcolor{blue}{L}}_\text{\color{black}{Dorm 5}}-
\underbrace{H-\textcolor{blue}{L}}_\text{\color{black}{Dorm 6}},$$

Each student ends up with exactly one roommate whose type always corresponds to the assigned treatment arm. The list is also flexible for larger dorms. For example, if dormitories carried four students, the dorms' composition would perfectly align with student-peer-type combinations:
$$\underbrace{\textcolor{blue}{H}-\textcolor{blue}{H}-\textcolor{blue}{H}-\textcolor{blue}{H}}_\text{\color{black}{Dorm 1}}-\underbrace{L-L-L-L}_\text{\color{black}{Dorm 2}}-\underbrace{H-\textcolor{blue}{L}-H-\textcolor{blue}{L}}_\text{\color{black}{Dorm 3}}$$

Noncompliance can nonetheless occur in dormitories of sizes that do not fully conform to student-peer-type combinations. For example, consider dorms with three students. The allocation would be as follows: 
$$\underbrace{\textcolor{blue}{H}-\textcolor{blue}{H}-\textcolor{blue}{H}}_\text{\color{black}{Dorm 1}}-\underbrace{\textcolor{blue}{H}-L-L}_\text{\color{black}{Dorm 2}}
-\underbrace{L-L-H}_\text{\color{black}{Dorm 3}}-\underbrace{\textcolor{blue}{L}-H-\textcolor{blue}{L}}_\text{\color{black}{Dorm 4}},$$

The list is also flexible enough to adapt to large dorms by placing, in the same bunk bed or in adjacent bunk beds, students who are adjacent to each other on the list. Overall, the list's order is directly linked to the physical proximity between two students in a dormitory. Students who are adjacent on the list are more likely to be near each other in the dormitories. In small dorms, the assigned peers likely share the same room. In bigger dorms, students and assigned peers are placed either in the same bunk bed or in beds next to each other. Our empirical strategy exploits the random variation on the list as a measure of proximity.

The specific design of the experiment in this paper follows the same structure but with two treatments: more socially central peers and higher-achieving peers. With four student types, there are ten student-peer type combinations. These combinations are randomly ordered on a list using the same guidelines as the ones described in the example above. In general, after conditioning on student's type, the list generates random variation in the physical proximity between two students.  

We verified the compliance of school authorities with the experiment by comparing the lists we sent to the schools with the final allocation to dormitories. We corroborate whether school authorities followed the protocols by estimating equation \ref{eq: proximity} on the likelihood that two students are neighbors in dormitories. For small dormitories (fewer than five students), neighbors are students in the same dorm (roommates). For larger dormitories (more than five students), neighbors are students in either the same or the adjacent bunk bed.

\begin{figure}[h!]
    \caption{Impact of Proximity on Being Neighbors in Dormitories}
    \begin{center}
    \scalebox{0.8}{\begin{tabular}{cc}
Panel A: Effect of Proximity    & Panel B: Interaction with First-years       \\
         \includegraphics[scale=0.6]{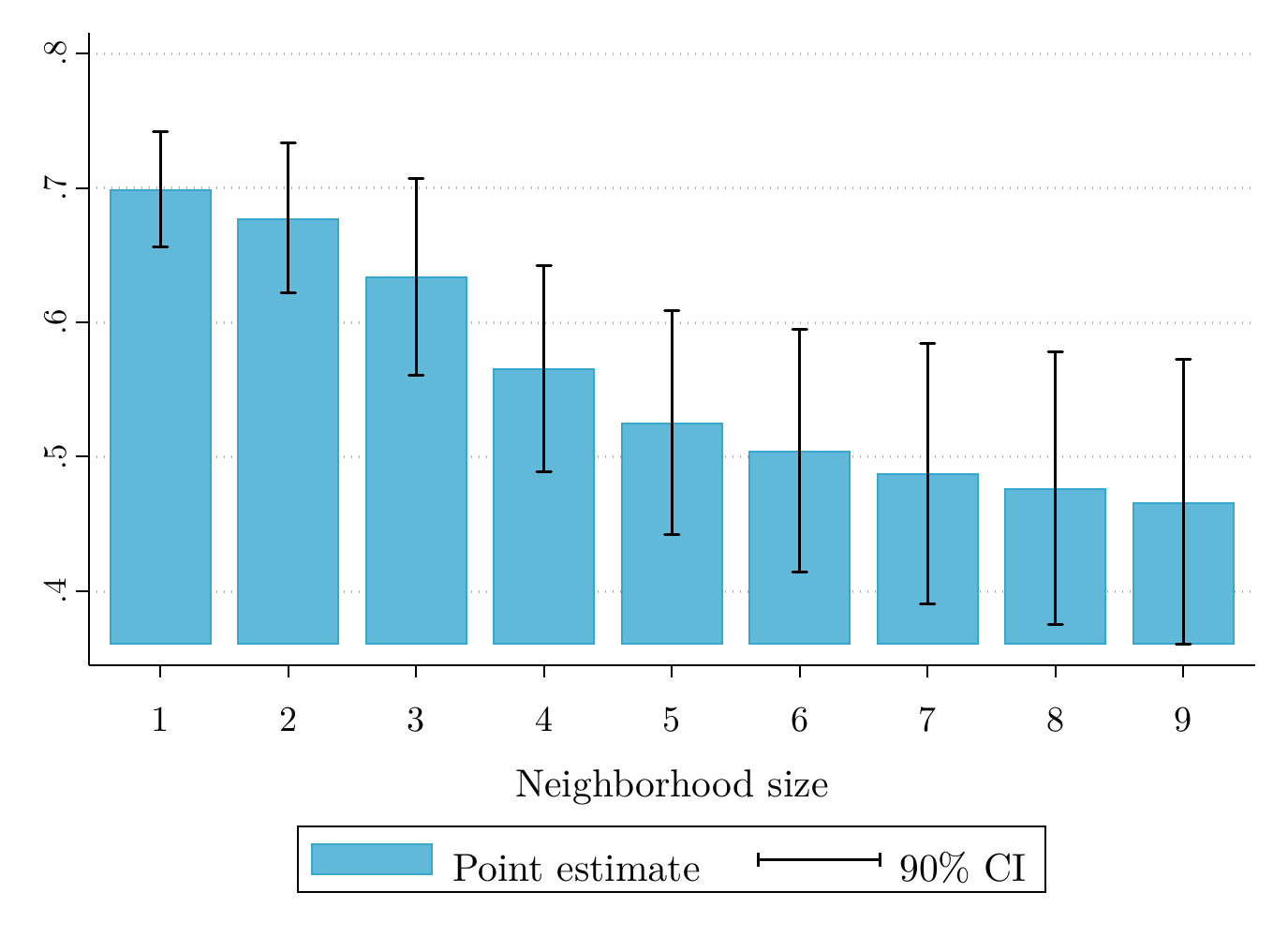} &
         \includegraphics[scale=0.6]{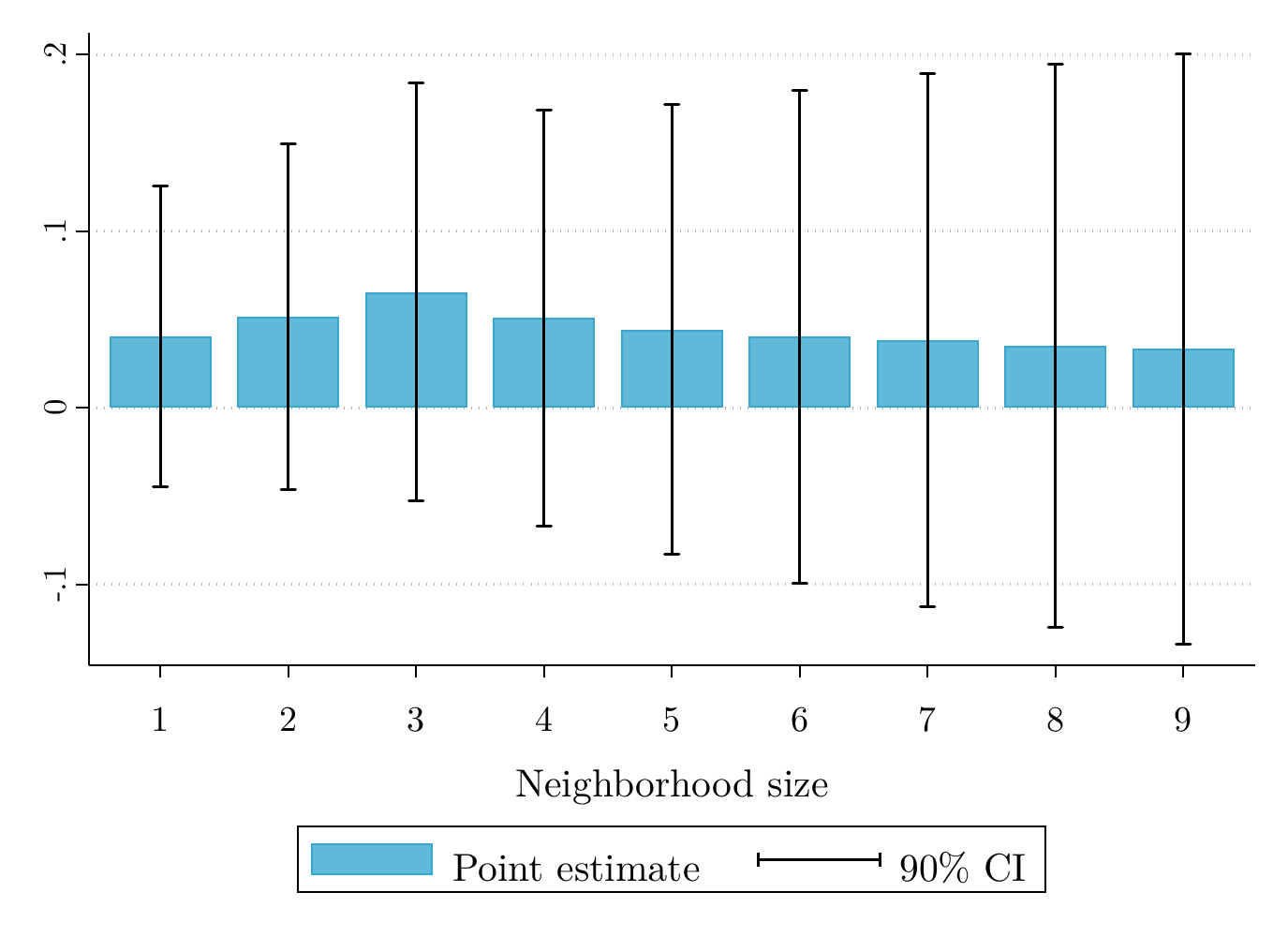} \\
    \end{tabular}}
    \end{center}
    \footnotesize{\textbf{Notes}: This figure presents estimates of Equation \eqref{eq: proximity} for different values of distance on the list, $d$, on the likelihood of being neighbors in a dormitory. Panel A shows the direct effect of being at a distance less or equal to $d$ on the list, and Panel B presents the differentiated effect for first-year students. \par}
    \label{fig: neighbors}
\end{figure}

Panel A of Figure \ref{fig: neighbors} presents the estimates for $d\in\lbrace 1, \ldots, 9 \rbrace$. The distance on the list predicts that students become neighbors in dormitories, which confirms that schools followed the list to assign students in the dormitories.  Panel B presents the interaction of the neighborhood's size dummies with being a first-year. There is no difference in how the distance on the list impacts the likelihood of being neighbors for first-years versus the other cohorts. 

\end{document}

%% file: tables/table2_tex.txt
Poor                &      -0.547** &               &       0.137   &      -0.664***&               &      -0.065   &      -0.474***&               &      -0.476***&      -0.046   \\
                    &     (0.169)   &               &     (0.085)   &     (0.113)   &               &     (0.113)   &     (0.106)   &               &     (0.143)   &     (0.075)   \\
Lower-achieving     &      -0.218   &               &       0.172** &      -0.360***&               &       1.056***&      -1.268***&               &      -0.070   &      -0.131*  \\
                    &     (0.155)   &               &     (0.077)   &     (0.105)   &               &     (0.101)   &     (0.097)   &               &     (0.131)   &     (0.071)   \\
Less central        &      -1.143***&               &      -0.270** &      -0.627***&               &      -0.443***&      -0.591***&               &       0.234** &      -1.072***\\
                    &     (0.247)   &               &     (0.098)   &     (0.158)   &               &     (0.121)   &     (0.137)   &               &     (0.101)   &     (0.163)   \\
mean                &       14.15   &               &        5.54   &        8.29   &               &        7.03   &        7.11   &               &        9.58   &        4.56   \\
N                   &       5,467   &               &       5,467   &       5,467   &               &       5,467   &       5,467   &               &       5,467   &       5,467   \\